\documentstyle[12pt]{article}
\begin{document}
\setlength{\textwidth}{6.125in}
\setlength{\textheight}{9.25in}
\newcommand{\bq}{\begin{equation}}
\newcommand{\eq}{\end{equation}}
\newcommand{\bqa}{\begin{eqnarray}}
\newcommand{\eqa}{\end{eqnarray}}
\newcommand{\nl}{\nonumber \\}
\newcommand{\mc}{Mon\-te Car\-lo}
\newcommand{\qmc}{Quasi-\mc}
\newcommand{\qran}{qua\-si-ran\-dom}
\newcommand{\suml}{\sum\limits}
\newcommand{\intl}{\int\limits}
\newcommand{\prodl}{\prod\limits}
\newcommand{\avg}[1]{\left\langle #1\right\rangle}
\newcommand{\avgf}[1]{\avg{#1}_f}
\newcommand{\si}{\sigma}

\begin{center}
{\Large {\bf Quasi-Monte Carlo, Discrepancies\\
and Error Estimates\footnote{presented at the 2nd International 
Conference on Monte Carlo 
and Quasi-Monte Carlo Methods in Scientific Computing, 
Salzburg, Austria, july 9-12,1996 }}}\\
\vspace{\baselineskip}
Jiri Hoogland\footnote{e-mail: t96@nikhefh.nikhef.nl,
 research supported by Stichting FOM.}\\
 NIKHEF, Amsterdam, The Netherlands\\
\vspace{\baselineskip}
Fred James\footnote{e-mail: james@mail.cern.ch}\\
CERN, Geneva, Switzerland\\
\vspace{\baselineskip}
Ronald Kleiss\footnote{e-mail: kleiss@sci.kun.nl,research supported
by Stichting FOM.}\\
University of Nijmegen, Nijmegen, The Netherlands\\
\end{center}\vspace*{\baselineskip}
\begin{center}{\bf Abstract}\end{center}
We discuss the problem of defining an estimate for the
error in \qmc\ integration. The key issue is the definition of an
ensemble of \qran\ point sets that, on the one hand, includes a sufficiency
of equivalent point sets, and on the other hand uses information on the
degree of uniformity of the point set actually used, in the
form of a discrepancy or diaphony. A few examples of such discrepancies
are given. We derive the distribution of our error estimate in the
limit of large number of points. In many cases, Gaussian central limits
are obtained. We also present numerical results
for the quadratic star-discrepancy for a number of \qran\ sequences.  

\section{The error problem}
We discuss the problem of integration of a square integrable
function over the $s$-dimensional unit hypercube $K=[0,1)^s$,
using a set of $N$ points $x_k$, $k=1,2,\ldots,N$. The actual
integral is $J=\int_K dx f(x)$, and its numerical estimate
is given by
\bq
S = {1\over N}\suml_{k=1}^N\;f(x_k)\;\;.
\eq
Depending on the way in which the points $x_k$ are chosen,
we distinguish different integration methods: if the points come
from some predetermined, deterministic scheme we have
a quadrature rule, if they are considered to be iid
uniform random numbers, we have \mc. An intermediate
case is that of \qmc, where the points are considered to be
part of a low-discrepancy sequence, but share the ergodic 
properties of a truly random sequence\footnote{We shall not
discuss the case of point sets with fixed, predetermined $N$.}.
The integration error is defined as $\eta\equiv S-J$. Good
integration methods are characterized by the fact that they
typically lead to a small value of $\eta$, but, more importantly,
allow one to obtain a good estimate of $\eta$. In the case of
\mc, $\eta$ is a stochastic variable, and hence has a probability
density $P(\eta)$. For \qran\ point sets used in \qmc, we may
(as we shall specify more precisely later on) also interpret
$\eta$ as having such a probability density: its form is the
subject of this contribution.\\

In true \mc, the error distribution $P(\eta)$ is obtained by
viewing the point set $\{x_1,x_2,\ldots,x_N\}$ as a typical
member of an ensemble of random point sets, governed by 
the obvious Cartesian combined probability density
\bq
P_N(x_1,x_2,\ldots,x_N) = 1\;\;,
\label{Cartesian}
\eq
so that the $x_k$ are indeed iid uniform random numbers. The error
$\eta$ is then a random variable over this ensemble, with the
following well-known results. In the first place, $S$
is an unbiased estimator of $J$ in the sense that $\avg{\eta}=0$,
where the average is over the ensemble of points sets. In the
second place, for large $N$, $P(N)$ approaches a normal
distribution according to the Central Limit Theorem. Finally,
the variance of this distribution is given by
$\avg{\eta^2} = \mbox{Var}(f)/N$, where Var denotes the variance.
Note that, since we average over the integration points, the
error distribution can depend only on the integrand itself.

The conceptual problem in the use of \qran\ rather than truly
random point sets is the following: a \qran\ point set is 
{\em not\/} a `typical' set of random points! 
Indeed, \qran\ point sets are very special, with carefully
built-in correlations between the various points so that each new point
tends to `fill a gap' left by the previous ones. The usual
\mc\ error estimate is therefore not really justified in
\qmc. On the other hand, many different error bounds
assure us that small errors are possible, and indeed likely
when we apply low-discrepancy point sets.
In the following, we shall discuss two approaches to a
solution of this conundrum. Obviously, we can only summarize
the main results here: technical details and pictures
can be found in the references.

\section{The Bayesian approach}
The first way around the aforementioned conceptual problem is
to interchange the r\^{o}les of integrand and point set: we
view the integrand $f(x)$ as a typical member of some underlying
class of functions and average over this class, so that the
error depends only on a property of the point set. In practice,
the choice of function class often entails a good deal of
idealism or pot luck, as usual in a Bayesian approach to
probability. We discuss several examples, in which we denote
by $\avgf{ }$ an average over the probability measure
governing the function class.

\subsection{The Wo\'{z}niakowski Lemma}
Let the integrand $f(x)$ be chosen according to the Wiener
(sheet) measure in $s$ dimensions. This measure is Gaussian, with
\bq
\avgf{f(x)} = 0\;\;\;,\;\;\
\avgf{f(x),f(y)} = \prodl_{\mu=1}^s\min(x^{\mu},y^{\mu})\;\;,
\eq
where the index $\mu$ labels the coordinates. We may then quote 
the Lemma from \cite{wozniakowski}: 
\bq
\avgf{\eta} = 0\;\;\;,\;\;\;
\avgf{\eta^2} = D_2(x_1^{\ast},x_2^{\ast},\ldots,x_M^{\ast})\;\;,
\eq
where $D_2$ stands for the $L_2$ norm of the well-known
star-dis\-crep\-an\-cy,
and the $x_k^{\ast}$ denotes the `reflected' point,
with $(x_k^{\ast})^{\mu} = 1-x^{\mu}$. In \cite{kleiss} it is shown,
moreover, that the distribution $P(\eta)$ in this case is a Gaussian.

We have here the interesting general fact that the choice of a particular
function class induces its own particular discrepancy. On the other
hand, in many cases (such as in particle physics phenomenology)
the Wiener measure is certainly not appropriate since it is dominated
by integrands that are not locally smooth. In \cite{woz2}, {\em folded\/}
Wiener sheet measures are studied with analogous results, but then again
these describe functions that are much too smooth.

\subsection{Induced discrepancies}
In \cite{kleiss}, we established the following general result. Let the
measure on the function class be such that 
\bq
\avgf{f(x_1)}=0\;\;,\;\;\avgf{f(x_1)f(x_2)} = \int dy\;h(x_1,y)h(x_2,y)
\eq
for all $x_{1,2}$ in $K$, for some $h(x,y)$. 
There is then an induced quadratic discrepancy, defined as follows:
\bq
\avgf{\eta^2} = \int dy\;g(y)^2\;\;\;,\;\;\;
g(y) = {1\over N}\suml_{k=1}^N h(x_k,y) 
- \intl_K dx\;h(x,y)\;\;.
\eq
Note that $h$ is not necessarily in the same function class as the $f$,    
and indeed $y$ may be defined in a space quite different from $K$. Note
that, whenever the function class measure is Gaussian, then
$P(\eta)$ will also be Gaussian. Generalizations to higher
moments can be found in \cite{kleiss,jiri}.

\subsection{Orthonormal function bases}
As an example in $s=1$, let $u_n(x)$ be an orthonormal set of functions
on $K$, as follows:
\bq
u_0(x) = 1\;\;\;,\;\;\;\intl_K dx\;u_m(x)u_n(x) = \delta_{m,n}\;\;,
\eq
with $m,n=0,1,2,\ldots$. Let $f(x)$ admit of a decomposition
\bq
f(x) = \suml_{n\ge0}v_n u_n(x)\;\;,
\eq
and choose the measure such that the $v_n$ are normally distributed
around zero, with variance $\si_n^2$. The induced quadratic
discrepancy $D_2^{orth}$ is then defined\footnote{This measure
of discrepancy is also called {\em diaphony}.} as
\bq
\avgf{\eta^2} = {1\over N}D_2^{orth}\;\;\;,\;\;\;
D_2^{orth} = {1\over N}\suml_{n>0}\si_n^2
\suml_{k,l=1}^N\;u_n(x_k)u_n(x_l)\;\;.
\eq
A special case is that of the Fourier class, which has
\bq
u_{2n} = \sqrt{2}\cos(2\pi nx)\;\;,\;\;
u_{2n-1} = \sqrt{2}\sin(2\pi nx)\;\;,\;\;n=1,2,3,\ldots
\eq
The physically reasonable requirement that the {\em phase\/} of
each mode $n$ (made up from $u_{2n-1}$ and $u_{2n}$) is uniformly
distributed forces us to have the Gaussian measure, with in 
addition $\si_{2n-1}=\si_{2n}$, which property corresponds to
translational invariance. We then have
\bqa
D_2^{orth} & = & {2\over N}\suml_{n>0}\si_{2n}^2
\left|\suml_{k=1}^N\exp(2i\pi nx_k)\right|^2\;\;.
\eqa
Obviously, other orthonormal function bases are also possible, such
as the system of Walsh functions;
a further discussion, including the straightforward generalization to
higher dimension, can be found in \cite{jiri,fred,hk1,hk2}. Note that
all such quadratic discrepancies are nonnegative by construction.

\section{The discrepancy-based approach}
Another way of establishing integration error estimates, which in our
opinion does more justice to the spirit of \mc, is the following. 
Instead of considering all point sets of $N$ truly random points,
with the Cartesian probability density (\ref{Cartesian}), we restrict
ourselves to those point sets that have a given value of discrepancy, for
some predefined type of discrepancy. In this way, information on
the discrepancy performance of one's favorite \qran\ number sequence
can be incorporated in the error estimate. 

\subsection{Non-Cartesian distribution of points}
We have then, instead of (\ref{Cartesian}), a combined
probability density $P_N$ for the $N$ points as follows. Let
$D_N(x_1,\ldots,x_N)$ be {\em some\/} discrepancy defined on sets
of $N$ points in $K$, and suppose its value for the actual
point set that is used in the integration be $w$. Then,
\bqa
H(w) & = & \intl_K\;dx_1\cdots dx_N\;
\delta\left(D_N(x_1,\ldots,x_N)-w\right)\;\;,\nl
P_N(w;x_1,\ldots,x_N) & = & {1\over H(w)}
\delta\left(D_N(x_1,\ldots,x_N)-w\right)\nl
& = & 1-{1\over N}F_N(w;x_1,\ldots,x_N)\;\;,
\eqa
so that $F_N$ measures the deviation from Cartesian (iid) uniformity.
The quantity $H(w)$ is of course just the probability density for
the discrepancy over the iid uniform random numbers, an object interesting
in its own right.
Let us also define marginal deviations as
\bq
F_k(w;x_1,\ldots,x_k) = \intl_K\;dx_{k+1}\cdots dx_N 
F_N(w;x_1,\ldots,x_N)\;\;.
\eq
We can then simply establish, for instance, that,
provided $F_1(w;x)$ vanishes for all $x$,
\bqa
\avg{\eta^2} & = & {1\over N}\left[
{\mbox{Var}}(f) - {N-1\over N}\intl_K\;dx dy\;f(x)f(y)
F_2(w;x,y)\right]\;\;,\nl
\avg{\eta} & = & 0\;\;,
\eqa
where $\avg{}$ now denotes averaging with respect to $P_N(w;.)$.
It is seen that we may expect a reduced error if $F_2$ is positive
when $x$ and $y$ are `close' together in some sense, {\it i.e.\/}
if the points in the point set `repel' each other. Note that only
a small, ${\cal O}(1/N)$, deviation from uniformity is sufficient.

\subsection{Error probability distribution}
In many cases,it is actually possible to compute the $F_2$ mentioned 
above. In fact, especially in the case of discrepancies defined using
orthonormal function bases, we can do much more. Using a Feynman-diagram
technique described in detail in \cite{hk1,hk3}, we can establish
results for $P(\eta)$ as an asymptotic expansion in $1/N$. To leading
order, we have
\bqa
P(\eta) & = & {\sqrt{N/2\pi}\over 2\pi i H(w)}
\intl_{-i\infty}^{+i\infty}dz{1\over\sqrt{B(z)}}
\exp\left[A(z)-{\eta^2N\over 2B(z)}\right]\;\;,\nl
A(z) & = & -wz -{1\over2}\suml_{n>0}\log(1-2z\si_n^2)\;\;,\nl
B(z) & = & \suml_{n>0}{v_n^2\over1-2z\si_n^2}\;\;,
\eqa
where the $z$ integral runs to the left of any singularities.
This result holds, for $N$ asymptotically large, for any 
discrepancy measure based on orthonormal functions as discussed above,
and, moreover, for any reasonable $f$, even if it is not in the
function class based on these orthonormal functions. The $1/N$ corrections
are fully calculable, although we have not done so yet. Two corollaries
follow immediately. In the first place, 
\bq
\intl_0^{\infty}dw\;H(w)P(\eta) = \sqrt{N/2\pi V}e^{-\eta^2N/2V}\;\;\;,
\;\;\;V\equiv\suml_{n>0}v_n^2 = \mbox{Var}(f)\;\;,
\eq
which recovers the Central Limit Theorem valid over the
whole ensemble of $N$-point point sets with any $w$. In the second place,
we obtain an integral representation for $H(w)$ by insisiting that
$P(\eta)$ be normalized to unity:
\bq
H(w) = {1\over2\pi i}\intl_{-i\infty}^{+i\infty}dz\;
\exp\left[-zw-{1\over2}\suml_{n>0}\log(1-2z\si_n^2)\right]\;\;.
\eq
Generalizations of these results only affect the sums over $n$.

\subsection{Application 1: equal strengths}
A simple model for a discrepancy is obtained by taking
$\si_n=1/2M$ for $n=1,2,\ldots,2M$, and zero otherwise (with trivial 
extension to more dimensions). Let us then decompose the variance of the
integrand as follows:
\bq
V = \mbox{Var}(f) = \suml_{n>0}v_n^2 = V_1 + V_2\;\;,\;\;
V_1=\suml_{n=1}^{2M}v_n^2\;\;,\;\;V_2=\suml_{n>2M}v_n^2\;\;,
\eq
so that $V_1$ contains that part of the variance to which the discrepancy
is sensitive (the `covered' part) and $V_2$ the rest (the `uncovered' part).
We then have
\bqa
H(w) & = & {M^M\over\Gamma(M)}w^{M-1}e^{-Mw}\;\;\sim
\;\;\exp\left(-{M\over2}(w-1)^2\right)\;\;,\nl
P(\eta) & \sim & \left({N\over2\pi(wV_1+V_2)}\right)^{1/2}
\exp\left(-{\eta^2N\over2(wV_1+v_2)}\right)\;\;,
\eqa
where the approximations are valid for large $M$. We see that
a new central limit theorem holds, where the variance of $f$ has been
modified so that its covered part is reduced by a factor $w$, according to
intuition.

\subsection{Application 2: harmonic model in one dimension}
Let us concentrate on the case $s=1$, and take $\si_{2n-1}=\si_{2n}=1/n$,
so that $f$ is, on the average, square integrable, but its derivative is not.
In that case we have, 
\bq
H(w) = \suml_{m>0}\;(-1)^{m-1}m^2e^{-wm^2/2}\;\;,
\eq
which is, apart from a trivial scaling, precisely the probability
density of the Kolmogorov-Smirnov statistic. This is somewhat surprising
since that statistic is based on the $L_{\infty}$ norm of the standard
star-discrepancy, a totally different object. In addition, we conjecture,
that for values of $w$ small compared to its expectation value
$\avg{w}=\pi^2/3$, we shall have
\bq
P(\eta) \propto \exp\left[-{\eta^2N\over2C}\right]\;\;\;,\;\;\;
C = \mbox{Var}(f){w\over\avg{w}}\;\;;
\eq
this would again indicate a reduction of the error estimate for
small $w$. To date, we have not yet been able to prove this assertion.

\subsection{The quadratic star-discrepancy}
We have also obtained some results for the quadratic form of the
standard star-discrepanc \cite{jiri,fred}. 
Although this is not based on orthonormal functions
and the analysis is hence more complicated, we have obtained
the moment-generating function $G(z)=\avg{e^{zw}}$
for asymptotically large $N$, where $w$ now stands for $N$ times the
quadratic star-discrepancy. More precisely, we have
\bqa
G(z) & = & \exp(\psi(z))/\sqrt{\chi(z)}\;\;,\nl
\psi(z) & = & -{1\over2}\suml_{n>0}Q_s(2n-1)\log(1-2z_n)\;\;,\nl
\chi(z) & = & {2^s\over2z}\suml_{n>0}Q_s(2n-1){z_n\over1-z_n}\;\;,\nl
z_n & = & (4/\pi^2)^s{2z\over(2n-1)^2}\;\;.
\eqa
Here, $Q_s(m)$ is the number of ways in which an odd positive integer $m$
can be written as a product of $s$ positive integers, including 1's.
The function $H(w)$ can now be computed numerically for different $s$
values. We have done so, and find that $H(w)$ very slowly approaches a
Gaussian distribution as $s$ increases. Indeed, the skewness of $H(w)$
is, for large $s$, approximately given by $(216/225)^{s/2}$ so that
the approach to normality is indeed slow.

\section{Numerical results for the quadratic star-discrepancy}
Since we now have $H(w)$ for the quadratic star-discrepancy, we can
reliably judge how well \qran\ number generators perform, since we can
compare the discrepancy of their output with the behaviour of truly
random points. Space does not permit us to show pictures, which can be
found in \cite{jiri,fred}. Here, we just describe the results.
We have computed the quadratic star-discrepancy for a good
pseudorandom generator (RANLUX), and for the Richtmeyer, Halton,
Sobol', and Niederreiter sequences. We did this both exactly, and
by Monte Carlo. The latter method is actually faster for $N$ larger 
than about 50,000 if we ask for 5\% accuracy. We made runs of up to
$N=150,000$, and considered the lowest and highest discrepancy value
in subsequent intervals of 1000. These we compared with the
expected value $(2^{-s}-3^{-s})/N$ for truly random points, and also
plotted the standardized\footnote{Again, here the $\avg{}$
stands for the expectation for truly random points.} 
form $\xi(w)=(w-\avg{w})/\sqrt{\mbox{Var}(w)}$.
We considered dimensions from 1 up to 20. In all dimensions, RANLUX
appears to mimic a truly random sequence quite well. The \qran\
generators perform very well in low dimensions, and generally the
discrepancy falls further and further below than of a random sequence as
$N$ increases. There are exceptions, however: for instance, the Sobol'
sequence for $s=11$ degrades, and is actually worse than random
for $N\sim60,000$, again rapidly improving for larger $N$. Apart from
taking this as a warning, we have not investigated the reason for such
behaviour in detail. The biggest surprise was when we plotted the
variable $\xi(w)$ (for instance, for $N=150,000$) as a function of
$s$. It appears that, as measured in this way, the performance of all
\qran\ generators {\em improves\/} with increasing $s$! For $s$ larger
than 15 or so, all generators become rather bad, which is of course due
to the fact that the onset of the asymptotic regime occurs for larger
$N$ as $s$ increases. But what is more striking is the fact that
the old-fashioned and simple 
Richtmyer generator performs as well as the modern,
sophisticated Sobol' and Niederreiter sequences. We take this as an
indication that the Richtmyer generator deserves more study, in particular
since we have not attempted any optimization of its `lattice point'.

\end{document}